\documentstyle[12pt,axodraw]{article}
\def\al{\alpha}
\def\be{\beta}

\def\de{\delta}
\def\ep{\epsilon}

\def\ka{\kappa}
\def\la{\lambda}

\def\si{\sigma}

\def\mn{{\mu\nu}}

\def\frac#1#2{{\textstyle{{#1}\over {#2}}}}

\def\lsim{\mathrel{\rlap{\lower4pt\hbox{\hskip1pt$\sim$}}
    \raise1pt\hbox{$<$}}}
\def\gsim{\mathrel{\rlap{\lower4pt\hbox{\hskip1pt$\sim$}}
    \raise1pt\hbox{$>$}}}
\def\sqr#1#2{{\vcenter{\vbox{\hrule height.#2pt
         \hbox{\vrule width.#2pt height#1pt \kern#1pt
         \vrule width.#2pt}
         \hrule height.#2pt}}}}

\newcommand{\beq}{\begin{equation}}
\newcommand{\eeq}{\end{equation}}
\newcommand{\bea}{\begin{eqnarray}}
\newcommand{\eea}{\end{eqnarray}}

\renewenvironment{thebibliography}[1]
 { \rm
   \begin{list}{\arabic{enumi}.}
    {\usecounter{enumi} \setlength{\parsep}{0pt}
     \setlength{\itemsep}{3pt} \settowidth{\labelwidth}{#1.}
     \sloppy
    }}{\end{list}}

\begin{document}
\titlepage


\vglue 1cm

\begin{center}
{{\bf One-Loop Renormalization of Pure Yang-Mills with Lorentz Violation \\}
\vglue 1.0cm
{Don Colladay and Pat McDonald\\}
\bigskip
{\it New College of Florida\\}
\medskip
{\it Sarasota, FL, 34243, U.S.A.\\}

\vglue 0.8cm
}
\vglue 0.3cm

\end{center}

{\rightskip=3pc\leftskip=3pc\noindent
The explicit one-loop renormalizability of pure Yang-Mills theory with Lorentz violation
is demonstrated.
The result is consistent with multiplicative renormalization as the required counter terms
are consistent with a single re-scaling of the Lorentz-violation parameters.
In addition, the resulting beta functions indicate that the CPT-even Lorentz-violating terms
increase with energy scale in opposition to the asymptotically free gauge coupling and
CPT-odd terms.
The calculations are performed at lowest-order in the Lorentz-violating terms as they are
assumed small.
}

\vskip 1 cm

\vskip 1 cm


\newpage

\baselineskip=20pt
{\bf \noindent I. INTRODUCTION}
\vglue 0.4cm

In practice, quantum field theories of interest are defined by
Lagrangians which are chosen to obey certain symmetry requirements.
It is well known that many such quantum field theories exhibit ultraviolet
divergences which arise due to the structure of the theory at small
distances.  These divergences can sometimes be removed by a singular
redefinition of the parameters defining the theory via the processes
of renormalization (the Standard Model affords an example).  It is
often the case that the calculations involved in establishing
renormalizability of a given theory involve a series of remarkable
cancellations in which multiple diagrams combine to produce precisely
the correct values of the counter-terms required for multiplicative
renormalization.  In these calculations, gauge symmetries play a
crucial role in obtaining the required cancellations.  This is
certainly the case in standard Yang-Mills theory \cite{ym}, where (internal)
Lorentz symmetries are included in the gauge group \cite{thooft}.  The
primary purpose of this paper is to investigate the explicit
renormalization properties of Yang-Mills theory in the presence of
Lorentz violation; more precisely, whether or not multiplicative
renormalization can be applied as in the conventional case.   As our
main result we establish explicit one-loop renormalizability
of pure Yang-Mills theory with Lorentz violation.

This work should be viewed as part of an extensive, systematic
investigation of Lorentz violation and its possible implications for
Planck-scale physics \cite{kps}.  Our calculations are carried out in
the framework of an explicit theory, the Standard Model Extension
(SME), which has been formulated to include possible Lorentz-violating
background couplings to Standard Model fields \cite{ck}.  This
formalism has also been extended to include gravity \cite{alangrav}.
Extensive calculations using the SME have led to numerous experiments
\cite{cpt98}, which have in turn placed stringent bounds on parameters
in the theory associated with electrons, photons \cite{kmewes}, and hadrons.  To
date, however, the same can not be said for nonabelian sectors of the
theory.  More precisely, there has been no systematic investigation of
the consistency of nonabelian gauge theories in the presence of
Lorentz violation and there are very few specific experimental bounds on the
nonabelian gauge parameters involved in the SME.

There are a number of reasons to investigate the consistency of
nonabelian gauge theories in the presence of Lorentz violation, and to
develop good bounds for the associated nonabelian gauge parameters.
For example, many new experimental results in the neutrino sector
suggest that some type of modification to the standard model will be
required to explain the results.  With this as motivation, the SME has
been applied to the neutrino sector \cite{kneut,oneut} with the hope
of generating a realistic model of oscillations.  Lorentz
violation may play an interesting role in weak interactions which contain gauge
bosons that have not yet been analyzed in detail.
In addition, QCD involves a strong coupling that
can be large and lead to more signigficant radiative corrections. 
Finally, there are also indications based on
general Renormalization Group analysis that gauge theories with
nonpolynomial interactions naturally tend to violate Lorentz
invariance \cite{kalt}.

Previous related theoretical work includes an analysis of the explicit
one-loop structure of Lorentz-violating QED and the resulting running
of the couplings \cite{klp}.  In this case, conventional
multiplicative renormalization is found to succeed and the beta
functions indicate a variety of running behaviors, all controlled by
the running of the charge (in contrast to QED, however, nonabelian
theories are generally asymptotically free \cite{gw}).  Part of this
analysis has been extended to allow for a curved-space background
\cite{bps}, while other analysis involved finite, but undetermined
radiative corrections due to CPT violation \cite{kafstuff}.  Other
related work includes a study of deformed instantons in the theory \cite{cm},
an analysis of the Casimir effect in the presence of Lorentz violation
\cite{ft}, and an analysis of gauge invariance of Lorentz-violating
QED at higher-orders \cite{ba}.  Some investigations into possible
Lorentz-violation induced from the ghost sector of
scalar QED have also been performed \cite{altghost}.

\vglue 0.6cm
{\bf \noindent I. NOTATION AND CONVENTIONS}
\vglue 0.4cm

The pure Yang-Mills lagrangian with Lorentz violation is taken as the
most general gauge invariant and power-counting renormalizable action
\bea
& & \hspace{-1cm} S(A) = - {1 \over 2} \int d^4 x Tr \left[ F^\mn F_\mn \right.
+  (k_F)_{\mn \alpha \beta} F^\mn F^{\al \be}
\nonumber \\
& & \hspace{3cm} + \left. (k_{AF})^\ka \ep_{\ka \la \mn}
(A^\la F^\mn - \frac 2 3 i g A^\la A^\mu A^\nu)\right]
\quad .
\label{minkac}
\eea
The generators of the Lie Algebra defined by $A^\mu = A^{a \mu} t^a$ are taken to satisfy
\beq
[t^a, t^b] = i f^{abc} t^c ~,
\eeq
where $f^{abc}$ are totally anti-symmetric structure constants.
The trace of a product of these generators is normalized to
\beq
Tr[t^a t^b] = C_2(r) \de^{ab}~,
\eeq
where $C_2(r)$ is the quadratic casimir of the representation $r$.
In the adjoint representation used for the gauge fields, this is written
$C_2(G)$.
The field tensor is defined as
\beq
F^\mn = - {i \over g}[D^\mu,D^\nu]~,
\eeq
where the covariant derivative is $D^\mu = \partial^\mu + i g A^\mu$.

The $k_F$ term is CPT even while the $k_{AF}$ term is CPT odd.
The radiative corrections can therefore be calculated separately.
In addition, properties of $F^{\mn}$ imply that the $k_F$ parameters have the
symmetries of the Riemann tensor and satisfy the Jacobi identity
\beq
k_F^{\la \mu\nu\rho} + k_F^{\la \nu\rho\mu} + k_F^{\la \rho\mu\nu} = 0 ~.
\eeq
To simplify the analysis, $k_F$ is taken to be trace free
\beq
g_{\mu\al} k_F^{\mn\al\be} = 0 ~.
\eeq
Nontrivial trace components are more easily handled using
coordinate redefinitions and can be removed from a pure Yang-Mills theory,
at least at the classical level \cite{cm}.
Expansion of the resulting Lagrangian and including a ghost field to integrate
over gauge yields the conventional terms
\bea
{\cal{L}}_{\rm LI} = & & - \frac 1 2 \left( \partial^\mu A^{a \nu} \partial_\mu A^a_\nu
- \partial^\nu A^{a \mu} \partial_\mu A^a_\nu + \frac 1 \xi (\partial^\mu A^a_\nu)^2
\right) + g f^{abc} A^{a \mu} A^{b \nu} \partial_\mu A^c_\nu
\nonumber \\
& & - \frac 1 4 g^2 f^{abe}f^{cde} A^{a\mu} A^{b \nu} A^c_\mu A^d_\nu + {\cal{L}}_{g}~,
\eea
where the ghost Lagrangian is written in terms of the scalar, anticommuting field $c$
\beq
{\cal{L}}_{g} = \overline c^a (-\partial^\mu \partial_\mu \de^{ac}
+ g \partial^\mu f^{abc}A^{b\mu}) c^c ~.
\eeq
The Lorentz-violating terms are separated into a CPT-even piece
\beq
{\cal L}_{\rm LVE} = (k_F)_{\mn\al\be}\left[ - \partial^\mu A^{a \nu} \partial^\al A^{a \be}
+ g f^{abc} (\partial^\mu A^{c \nu}) A^{a \al} A^{b \be}
- \frac 1 4 g^2 f^{abe} f^{cde} A^{a\mu} A^{b \nu} A^{c \al} A^{d \be} \right] ~,
\label{cpteven}
\eeq
and a CPT-odd piece
\beq
{\cal L}_{\rm LVO} = - \frac 1 2 (k_{AF})^\ka \ep_{\ka\la\mu\nu}A^{a \la}\partial^\mu A^{a\nu}
+ \frac 1 6 g (k_{AF})^\ka \ep_{\ka\la\mn} f^{abc} A^{a\la}A^{b\mu}A^{c\nu}~.
\label{cptodd}
\eeq
There will be a corresponding counter-term $\de_i$ for each of these terms in the above Lagrangian.
Specifically, the standard counter-terms are defined as
\bea
{\cal L}^{\rm ct}_{\rm LI} = & & - \frac 1 2 \de_3 \left( \partial^\mu A^{a \nu} \partial_\mu A^a_\nu
- \partial^\nu A^{a \mu} \partial_\mu A^a_\nu \right) + g \de_1^{3g} f^{abc} A^{a \mu} A^{b \nu} \partial_\mu A^c_\nu
\nonumber \\
& & - \frac 1 4 g^2 \de_1^{4g} f^{abe}f^{cde} A^{a\mu} A^{b \nu} A^c_\mu A^d_\nu
+ \de {\cal{L}}_g ~,
\eea
where $\de {\cal{L}}_g$ is the ghost counter-term.
In the present analysis, the ghost counter-term will turn out to be the same as in the conventional case
and will not be discussed further.
The CPT-even counter-terms are defined as
\bea
{\cal L}^{\rm ct}_{\rm LVE} = & &
(k_F)_{\mn\al\be}\left[ - \de_{k_F}^{2g}\partial^\mu A^{a \nu} \partial^\al A^{a \be}
+ g \de_{k_F}^{3g} f^{abc} (\partial^\mu A^{c \nu}) A^{a \al} A^{b \be} \right.
\nonumber \\
& & \left. - \frac 1 4 g^2 \de_{k_F}^{4g} f^{abe} f^{cde} A^{a\mu} A^{b \nu} A^{c \al} A^{d \be} \right]~.
\eea
while the CPT-odd ones are defined as
\beq
{\cal L}^{\rm ct}_{LVO} =
- \frac 1 2 (k_{AF})^\ka \ep_{\ka\la\mu\nu} \left[ \de_{k_{AF}}^{2g} A^{a \la}\partial^\mu A^{a\nu}
- \frac 1 3 g \de_{k_{AF}}^{3g} f^{abc} A^{a\la}A^{b\mu}A^{c\nu}\right]~.
\eeq
Explicit multiplicative renormalizability implies relations between the counter-terms
analogous to the standard case.
This occurs because the bare fields and couplings are defined in terms of the finite renormalized fields
and couplings as follows:
$A_B^\mu = \sqrt{Z_3} A^\mu$, $g_B = Z_g g$, $(k_F)_B = Z_{k_F} k_F$, and $(k_{AF})_B = Z_{k_{AF}}k_{AF}$.
Success of this prescription requires the appropriate counter-term relations that will be demonstrated
in this paper.
The calculation will be performed first for the CPT-even terms, then for the
CPT-odd terms.
Only first-order terms in Lorentz violation will be retained, as they are assumed to
be miniscule in magnitude.
It is then possible to consider the CPT-even and CPT-odd independently since they do not
mix (at lowest-order) under radiative corrections due to their different symmetry properties.

Throughout the paper, dimensional regularization is used to define the divergent
integrals and standard field-theoretic techniques are used to extract the divergent
terms \cite{ps}.

\vglue 0.6cm
{\bf \noindent II. FEYNMAN RULES FOR CPT-EVEN TERMS}
\vglue 0.4cm

The standard gluon propagator (in arbitrary $\xi$ gauge) is
\beq
\mu, a
\SetScale{0.3}
\begin{picture}(50,30)(0,13)
\Gluon(20,50)(160,50){5}{7}
\end{picture}
\nu, b
= {-i \de^{ab}}(g^\mn - (1 - \xi)l^\mu l^\nu)/l^2
\eeq
A single insertion of $k_F$ is indicated by a dot on the gluon line
\beq
\mu, a
\SetScale{0.3}
\begin{picture}(50,30)(0,13)
\Gluon(20,50)(160,50){5}{7}\Vertex(90,50){8}
\end{picture}
\nu, b
= -2i \delta^{ab} k_F^{\al\mu\be\nu}q_\al q_\be
\eeq
The standard three-point vertex is given by
\beq
\SetScale{0.3}
\begin{picture}(70,30)(0,13)
\Gluon(20,20)(72,50){5}{4}
\Gluon(72,110)(72,50){5}{4}
\Gluon(72,50)(124,20){5}{4}
\end{picture}
\hspace{-20pt}
= - g f^{abc}[g^\mn (k - p)^\rho + g^{\nu\rho}(p - q)^\mu + g^{\rho\mu}(q - k)^\nu]
\eeq
A correction to this vertex due to the $k_F$ term is denoted using a dot
\beq
\SetScale{0.3}
\begin{picture}(70,30)(0,13)
\Gluon(20,20)(72,50){5}{4}
\Gluon(72,110)(72,50){5}{4}
\Gluon(72,50)(124,20){5}{4}
\Vertex(72,50){8}
\end{picture}
\hspace{-20pt}
= 2 g f^{abc} [k_F^{\al\mn\rho}k_\al + k_F^{\al \nu\rho\mu}p_\al + k_F^{\al\rho\mu\nu} q_\al]
= 2g(V_k)^{abc}_{\mn\rho}
\eeq
In these expressions, the gluons are labeled using the associations $(a,\mu, k)$, $(b,\nu, p)$,
and $(c,\rho,q)$.
All momenta are defined going into the vertex.
The corresponding four-point vertices are given by
\beq
\SetScale{0.3}
\begin{picture}(70,30)(0,13)
\Gluon(30,8)(72,50){5}{4}
\Gluon(72,50)(114,92){5}{4}
\Gluon(30,92)(72,50){5}{4}
\Gluon(72,50)(114,8){5}{4}
\end{picture}
\hspace{-20pt}
= -i g^2 [f^{abe}f^{cde}(g^{\mu\rho} g^{\nu\si} - g^{\mu\si} g^{\nu\rho})+{\rm perms}]~.
\eeq
The gluons are labeled clockwise using $(a,\mu)$, $(b,\nu)$, $(c,\rho)$, and $(d,\si)$
and the two permutation terms are obtained from the first term by the
replacements $(b,\nu)\leftrightarrow (c,\rho)$ and $(b,\nu)\leftrightarrow(d,\si)$.
The correction to the four-point vertex is
\beq
\SetScale{0.3}
\begin{picture}(70,30)(0,13)
\Gluon(30,8)(72,50){5}{4}
\Gluon(72,50)(114,92){5}{4}
\Gluon(30,92)(72,50){5}{4}
\Gluon(72,50)(114,8){5}{4}
\Vertex(72,50){8}
\end{picture}
\hspace{-20pt}
= -2 i g^2[f^{abe}f^{cde}k_F^{\mn\rho\si} + {\rm perms}] = -2 i g^2 (V_k)^{abcd}_{\mn\rho\si}~,
\eeq
with two permutation terms given by the same replacements indicated in the standard
four-point vertex.

The standard divergent piece of the gluon two-point function is given in our notation by
\beq
\vspace{-10pt} \hfill \\
\SetScale{0.3}
\frac 1 2
\begin{picture}(70,30)(0,13)
\GlueArc(120,50)(40,0,180){5}{6}
\GlueArc(120,50)(40,180,360){5}{6}
\Gluon(20,50)(80,50){5}{3}
\Gluon(160,50)(220,50){5}{3}
\end{picture} +
{\rm tadpole} +
\begin{picture}(70,30)(0,13)
\DashArrowArc(120,50)(40,0,180){5}
\DashArrowArc(120,50)(40,180,360){5}
\Gluon(20,50)(80,50){5}{3}
\Gluon(160,50)(220,50){5}{3}
\end{picture}
= (\frac 5 3 + \frac 1 2(1 - \xi)) i(q^2 g^\mn - q^\mu q^\nu)\delta^{ab}\tilde g^2
\vspace{10pt} \hfill \\
\eeq
where the $1/2$ denotes the symmetry factor, the dotted lines indicate ghosts,
\beq
\tilde g^2 = {g^2 \over (4 \pi)^2}C_2(G)\Gamma(2 - \frac d 2)~,
\eeq
and $C_2(G)$ is the standard quadratic Casimir element of the adjoint
representation of the algebra.
The required counter-term is therefore
\beq
\de_3 = (\frac 5 3 + \frac 1 2(1 - \xi)) \tilde g^2 = Z_3 - 1
\eeq
where $A^\mu_B = \sqrt{Z_3}A^\mu$ gives the field strength renormalization.

Explicit calculation of the corresponding standard three- and four-point functions lead to the
following standard counter-terms
\beq
\de_1^{3g} = (\frac 2 3 + \frac 3 4 (1 - \xi) \tilde g^2 = Z_g Z_3^{3/2} - 1
\eeq
where $g_B = Z_g g$ is the charge renormalization, and
\beq
\de_1^{4g} = - (\frac 1 3 + (1 - \xi)) \tilde g^2 = (Z_g Z_3)^2 - 1~.
\eeq
All of these results are consistent with multiplicative renormalization
conditions $Z_g = 1 - (11/6)\tilde g^2$ and $Z_3 = 1 + (5/3 + 1/2(1 - \xi))\tilde g^2$.
Note that the gauge contribution $\xi \ne 1$ is absorbed entirely by $Z_3$.
This will turn out to be true in the Lorentz-violating case as well.

\vglue 0.6cm
{\bf \noindent III. ONE-LOOP CPT-EVEN RESULTS}
\vglue 0.4cm

Incorporation of the Lorentz-violating terms in Eq.(\ref{cpteven}) yields three
topologically distinct and nontrivial correction terms to the two-point function.
The direct corrections to the three-point vertices within the two-point function
yields
\beq
\vspace{-10pt} \hfill \\
\frac 1 2
\SetScale{0.3}
\begin{picture}(70,30)(0,13)
\GlueArc(120,50)(40,0,180){5}{6}
\GlueArc(120,50)(40,180,360){5}{6}
\Gluon(20,50)(80,50){5}{3} \Vertex(80,50){8}
\Gluon(160,50)(220,50){5}{3}
\end{picture} + \frac 1 2
\begin{picture}(70,30)(0,13)
\GlueArc(120,50)(40,0,180){5}{6}
\GlueArc(120,50)(40,180,360){5}{6}
\Gluon(20,50)(80,50){5}{3}
\Gluon(160,50)(220,50){5}{3} \Vertex(160,50){8}
\end{picture}
= (\frac {28} 3 - (1 - \xi)) i \de^{ab}\tilde g^2 k_F^{\al\mu\be\nu}q_\al q_\be~.
\label{2pta}
\eeq

There is one more nontrivial diagram that includes an insertion on an internal line
\beq
\vspace{-10pt} \hfill \\
\SetScale{0.3}
\begin{picture}(70,30)(0,13)
\GlueArc(120,50)(40,0,180){5}{6}
\GlueArc(120,50)(40,180,360){5}{6}
\Gluon(20,50)(80,50){5}{3} \Vertex(120,90){8}
\Gluon(160,50)(220,50){5}{3}
\end{picture}
=  (- \frac 4 3 + 2(1 - \xi)) i \de^{ab}\tilde g^2 k_F^{\al\mu\be\nu}q_\al q_\be.
\vspace{10pt} \hfill \\
\label{2ptb}
\eeq
Note that there is only one diagram of this type as an insertion on the other
internal line is not topologically distinct.
Combining the above diagrams yields
the required two-point counter-term
\beq
\delta_{k_F}^{2g} = (4 + \frac 1 2 (1 - \xi)) \tilde g^2 ~ ,
\eeq
to cancel the divergent piece.
This indicates that the bare couplings $(k_F)_B$ should be renormalized using the factor
$(k_F)_{B\mu\nu\rho\si} = Z_{k_F} (k_F)_{\mn\rho\si}$ with
\beq
Z_{k_F} = 1 + \frac 7 3 \tilde g^2~.
\label{kfrenorm}
\eeq
Note that the the renormalization of the physical couplings $k_F$ and $g$ remain independent of the
gauge choice.
Re-scaling $k_F$ in this way has implications for the required values for the Lorentz-violating three- and four-point
vertices that will now be analyzed.

Calculation of the corrections to the three-point vertex yields the following
results

\beq
\SetScale{0.3}
\vspace{-10pt} \hfill \\
\begin{picture}(50,30)(0,13)
\Gluon(-15,0)(37,30){5}{4}
\Gluon(72,150)(72,90){5}{4}
\Gluon(107,30)(159,0){5}{4}
\GlueArc(72,50)(40,90,210){5}{4}
\GlueArc(72,50)(40,210,330){5}{4}
\GlueArc(72,50)(40,-30,90){5}{4}
\Vertex(37,30){8}
\Vertex(72,90){8}
\Vertex(107,30){8}
\end{picture}
= (3-3(1 - \xi)) g \tilde g^2 (V_k)^{abc}_{\mn\rho}~ ,
\vspace{10pt} \hfill \\
\eeq

\beq
\SetScale{0.3}
\vspace{-10pt} \hfill \\
\begin{picture}(50,30)(0,13)
\Gluon(-15,0)(37,30){5}{4}
\Gluon(72,150)(72,90){5}{4}
\Gluon(107,30)(159,0){5}{4}
\GlueArc(72,50)(40,90,210){5}{4}
\GlueArc(72,50)(40,210,330){5}{4}
\GlueArc(72,50)(40,-30,90){5}{4}
\Vertex(37,70){8}
\Vertex(72,10){8}
\Vertex(107,70){8}
\end{picture}
= (- \frac 3 4 + \frac 3 4 (1 - \xi)) g \tilde g^2 (V_k)^{abc}_{\mn\rho}~ ,
\vspace{10pt} \hfill \\
\eeq

\beq
\SetScale{0.3}
\vspace{-10pt} \hfill \\
\frac 1 2
\begin{picture}(50,30)(0,13)
\Gluon(30,-32)(72,10){5}{4}
\Gluon(72,150)(72,90){5}{4}
\Gluon(114,-32)(72,10){5}{4}
\GlueArc(72,50)(40,90,270){5}{6}
\GlueArc(72,50)(40,-90,90){5}{6}
\Vertex(72,90){8}
\Vertex(72,10){8}
\end{picture} + {\rm cross~terms}
 = (-9 + \frac 9 4 (1 - \xi)) g \tilde g^2 (V_k)^{abc}_{\mn\rho}~ ,
\vspace{20pt} \hfill \\
\eeq

\beq
\SetScale{0.3}
\vspace{-10pt} \hfill \\
\begin{picture}(50,30)(0,13)
\Gluon(30,-32)(72,10){5}{4}
\Gluon(72,150)(72,90){5}{4}
\Gluon(114,-32)(72,10){5}{4}
\GlueArc(72,50)(40,90,270){5}{6}
\GlueArc(72,50)(40,-90,90){5}{6}
\Vertex(32,50){8}
\end{picture} + {\rm cross~terms}
= (\frac 3 4 - \frac 3 2(1 - \xi)) g \tilde g^2 (V_k)^{abc}_{\mn\rho}~ .
\vspace{20pt} \hfill \\
\eeq
In the above diagrams, symmetry factors of $1/2$ and appropriate
cross-terms are indicated.
The presence of several dots in a single diagram indicates a sum
over diagrams with each choice for the vertex or internal line correction.
The sum of these diagrams yields a required
counter term of
\beq
\de_{k_F}^{3g} = (3 + \frac 3 4(1 - \xi)) \tilde g^2 ~.
\eeq
This is in fact perfectly consistent with multiplicative renormalization condition
of equation (\ref{kfrenorm}) predicted by the correction to the two-point
function.  Specifically, the above counter-term takes the form $\de_k^{3g} = Z_g Z_3^{3/2} Z_{k_F} - 1$
as expected.

An analogous calculation of the four-point function (performed in
$\xi=1$ gauge for calculational simplicity)
yields the counter-term
\beq
\de_k^{4g} = 2 \tilde g^2 ~,
\eeq
which is again consistent with the multiplicative renormalization prediction
of Eq.(\ref{kfrenorm})
$\de_{k_F}^{4g} = Z_g^2 Z_3^2 Z_{k_F} - 1$.

\vglue 0.6cm
{\bf \noindent IV. ONE-LOOP CPT-ODD RESULTS}
\vglue 0.4cm

Next, attention is turned to the CPT-odd $k_{AF}$ terms given in Eq.(\ref{cptodd}).
The Feynman rules for the quadratic terms (indicated again by corresponding dots) are given by
\beq
\mu, a
\SetScale{0.3}
\SetScale{0.3}
\begin{picture}(50,30)(0,13)
\Gluon(20,50)(160,50){5}{7}\Vertex(90,50){8}
\end{picture}
\nu, b
= \delta^{ab} (k_{AF})^\ka \ep_{\ka\mu\be\nu}p^\be~,
\eeq
where the momentum $p$ flows from $\mu$ to $\nu$.
The cubic term contributes the following correction to the three-point
vertex
\beq
\SetScale{0.3}
\begin{picture}(70,30)(0,13)
\Gluon(20,20)(72,50){5}{4}
\Gluon(72,110)(72,50){5}{4}
\Gluon(72,50)(124,20){5}{4}
\Vertex(72,50){8}
\end{picture}
= i g f^{abc} (k_{AF})^\ka \ep_{\ka\mu\nu\rho} ~.
\eeq
Note that $k_{AF}$ has dimensions of energy and therefore decreases
the power-counting divergences of all graphs by one power.

The nontrivial contributions of the CPT-odd terms to the two-point function are
the same as shown in the diagrams of Eq.(\ref{2pta}) and Eq.(\ref{2ptb}).
The resulting required counter-term is given by
\beq
\de_{k_{AF}}^{2g} = (-2 + \frac 1 2(1 - \xi)) \tilde g^2~.
\eeq
This suggests renormalization of the $k_{AF}$ parameter according to
$(k_{AF})_B^\ka = Z_{k_{AF}}(k_{AF})^\ka$ with
\beq
Z_{k_{AF}} = 1 - \frac {11} {3} \tilde g^2~.
\label{kafrenorm}
\eeq
Note that the renormalization of the physical coupling is independent of gauge.
This again has implications for the radiative corrections to the three-point function.
An explicit calculation analogous to the one involving the CPT-even terms yields the
required counter-term
\beq
\de_{k_{AF}}^{3g} = (-3 + \frac 3 4(1 - \xi)) \tilde g^2~.
\eeq
This is exactly equal to the
prediction based on the multiplicative renormalization structure
$\de_{k_{AF}} = 1 + Z_{k_{AF}} Z_g Z_3^{3/2}$.
Note that there is no divergent contribution to the four-point diagram
due to the dimensionality of $k_{AF}$.

\vglue 0.6 cm
{\bf \noindent V. BETA FUNCTIONS}
\vglue 0.4 cm
Tacitly assuming for the moment that our renormalization prescription
can be extended to all orders,
the renormalization constants $Z_{k_F}$ and $Z_{k_{AF}}$ can be used to deduce
the one-loop beta functions for these parameters.
Following the developments presented in \cite{klp}, use is made of
\beq
\beta_{x_j} = \lim_{\ep \rightarrow 0}\left[ - \rho_{x_j} a_1^j
+ \sum_{k=1}^N \rho_{x_k} x_k {\partial a_1^j \over \partial x_k}\right]~,
\eeq
where $x_j$ represents an arbitrary running coupling in the theory.
The parameters $\rho_{x_j}$ are determined by comparing the mass dimension
of the renormalized parameters to the bare parameters in $d = 4 - 2 \ep$ dimensions
\beq
x_{jB} = \mu^{\rho_{x_j}\ep}Z_{x_j}x_j~.
\eeq
This gives the values
\beq
\rho_g = 1~, \quad \rho_{k_F} = \rho_{k_{AF}} = 0~.
\eeq
As in the QED case \cite{klp}, the coupling $g$ completely controls the running of the
Lorentz-violating parameters.
The resulting beta functions are
\beq
\beta_g = -{11 g^3 \over 3(4 \pi)^2}C_2(G) ~,
\eeq
the same as the conventional case, and
\beq
\beta_{k_F} = {14 g^2 \over 3 (4 \pi)^2}C_2(G) k_F ~,
\quad
\beta_{k_{AF}} = - {22 \over 3} {g^2 \over (4 \pi)^2}C_2(G) k_{AF}~.
\eeq
Note that the Lorentz indices have been suppressed for simplicity.
Introducing the parameter
\beq
Q(\mu) = 1 + {22 g_0^2 \over 3(4 \pi)^2}C_2(G)\ln{\mu \over \mu_0}~,
\eeq
and defining the initial conditions $x_{j0} = x_j(\mu_0)$ at the scale $\mu_0$
allows the solutions to the renormalization group equations to be put in the form
\beq
g^2(\mu) = Q^{-1} g_0^2~,
\eeq
reproducing the conventional result of asymptotic freedom for $g$
and
\beq
k_F = (k_F)_0 Q^{7/11} ~, \quad k_{AF} = (k_{AF})_0 Q^{-1}~.
\eeq
The CPT-odd $k_{AF}$ term behaves analogously to $g$, while the CPT-even
parameter $k_F$ increases with energy scale.
This suggests that Lorentz-violation involving CPT-even effects in the strong
interactions (where $g_0$ can be relatively large) may increase significantly at higher-energy scales.

\vglue 0.6 cm
{\bf \noindent V. SUMMARY}
\vglue 0.4 cm

In short, multiplicative renormalization is remarkably successful for
the case of Yang-Mills theory with Lorentz violation.
There are many opportunities for inconsistencies to arise, but they
do not actually occur, at least at the one-loop level.
The main numerical results can be summarized by the values for the renormalization
factors given in Eqs.(\ref{kfrenorm}) and (\ref{kafrenorm}).
These results yield beta functions that can be used to determine the
running of the Lorentz-violating couplings under the assumption
of full renormalizabilty of the model.
This indicates that the $k_F$ parameter increases with energy scale,
while $k_{AF}$ decreases with energy scale.  The behavior of the standard
coupling $g$ remains asymptotically free, as in the conventional case.
Note that the beta functions are independent of the gauge parameter $\xi$, in accord
with the results found in \cite{klp}.
Our results indicate that
in contrast to the QED case, the couplings in QCD may be significantly
larger and lead to observable effects.
The running of the parameters may have more of an impact due to the advantage
gained by the size of $\alpha_s$ compared to $\alpha$.
A detailed study will involve extending the current model to include the
quark sector.

Note that diagrams involving ghosts did not appear in any of the calculations
of Lorentz-violating corrections.
This can be attributed to the fact that ghost sector Lorentz violation terms can only
have symmetry properties that match the trace components of $k_F$
resulting in no contribution (at lowest-order) to trace-free Lorentz-violation parameters.
Since the trace components of $k_F$ were neglected in this paper, the ghosts only contribute
to conventional, Lorentz invariant terms and do not need modification.
The required counter-term structure in the ghost sector is therefore the same as in
the conventional case.
As noted previously, for a pure Yang-Mills theory, trace components of $k_F$ are
better handled by transforming to skewed coordinates with a nondiagonal metric.
Violations in the ghost sector should naturally arise due to the resulting skewed form of
the metric.

\vglue 0.6cm
{\bf\noindent REFERENCES}
\vglue 0.4cm

\end{document}